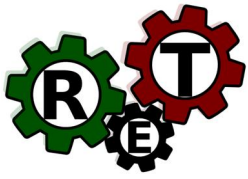

WORKSHOP SUMMARY
1st International Workshop on Requirements and Testing (RET '14)
Co-located with IEEE International Requirements Engineering Conference (RE 2014),
August 26th, 2014, Karlskrona, Sweden

Michael Felderer
Elizabeth Bjarnason
Markus Borg
Michael Unterkalmsteiner
Mirko Morandini
Matt Staats

# 1  Objectives

The main objective of the RET workshop was to explore the interaction of Requirements Engineering (RE) and Testing, i.e. RET, in research and industry, and the challenges that result from this interaction. While much work has been done in the respective fields of requirements engineering and testing, there exists much more than can be done to understand the connection between the processes of RE and of testing. RET will provided a forum for exchanging ideas and best practices for aligning RE and testing, and in particular intended to foster industry-academia collaboration on this topic. Towards this, RET invited submissions exploring how to best align RE and testing including processes, practices, artefacts, methods, techniques and tools. Submissions on softer aspects like the communication between roles in the engineering process were also welcomed.

RET accepted technical papers with a maximal length of 8 pages presenting research results or industrial practices related to the coordination of Requirements Engineering and Testing, as well as position papers with a minimal length of 2 pages introducing challenges, visions, positions or preliminary results within the scope of the workshop. Experience reports and papers on open challenges in industry were especially welcome.

# 2  Organization

The 1st International Workshop on Requirements and Testing (RET'14) was held on August 26, 2014 and was co-located with the IEEE International Requirements Engineering Conference (RE 2014). The website of the workshop is available online at http://webhotel.bth.se/re14/ret/. The workshop was organized by Michael Felderer (University of Innsbruck) as a general chair, Elizabeth Bjarnason (Lund University) and Matt Staats (Google Zurich) as program co-chairs, as well as Mirko Morandini (University of Trento), Michael Unterkalmsteiner (Blekinge Institute of Technology), and Markus Borg (Lund University) as co-chairs.

# 3  Program Summary

The program of RET'14 comprised an introductory part, three paper presentation sessions and a mapping exercise. Besides a welcome note presenting the objectives of the workshop and a presentation of the workshop sponsor EASE - The Industrial Excellence

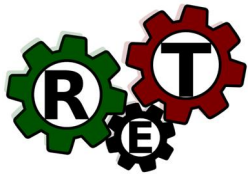

Centre for Embedded Applications Software Engineering (http://ease.cs.lth.se/), Magnus Ohlsson gave an invited talk "The Agile Hangover – Handling Testable Agile Requirements". In his talk, especially the role of communication of testers and other stakeholders and roles is highlighted for successful problem-oriented acceptance and solution-oriented system testing. It is especially required to integrate testers themselves (not only test managers) into problem and solution related discussions.

The first paper presentation session comprised four papers on RET challenges and practices. The talk "On the Delicate Balance between RE and Testing: Experiences from a Large Company" presents experiences from a large company which significantly downplayed requirements engineering activities while shifting the focus to testing and QA activities. A critical success factor was that the QA and testing staff got early involved into the requirements activities and discussions and helped to detail them. The talk "Revisiting the Challenges in Aligning RE and V&V: Experiences from the Public Sector" confirms that most challenges found in other studies also apply to the public sector, including the challenges of aligning goals within an organization, specifying high-quality requirements, and verifying quality aspects. The talk "Testers Learning Requirements" provides insights from practice on problems of the low integration of testers into requirements engineering activities and argues that integration of testers into requirements creation and review would be valuable. It is also highlighted that better training of testers in RE and requirements engineers in testing could help to improve the situation. In the talk "A/B Testing: A Promising Tool for Customer Value Evaluation" A/B testing is proposed to complement qualitative user research and to offer a potential way to validate the value which system improvements bring to customers. In the final panel discussion of this session it was highlighted that a holistic view of requirements engineering and testing is needed and that especially soft aspects like communication and education are critical success factors for the interaction of RE and testing. Furthermore, it is highlighted that the relationship of RE and testing in context of open source software development becomes more important also for established businesses and requires additional investigation.

The second paper presentation session comprised three papers on quality requirements. In the talk "Verifying Security Requirements using Model Checking Technique for UML-Based Requirements Specification" it is stressed that it is difficult to correctly specify adequate security requirements during the initial phases of the software development process. For this purpose, a method is proposed to verify security requirements specified on the basis of the Unified Modelling Language using model checking and Common Criteria security knowledge. In the talk "Using Automated Tests for Communicating and Verifying Non-functional Requirements" an alternative way to communicate and verify non-functional requirements, i.e., to add them to the tool-chain as automated tests and checkers, is proposed. This provides developers with fast automated feedback when they do mistakes. Practical experiences indicate that the productivity increases and the number of non-compliant non-functional requirements lowers when using the tool-chain feedback instead of using guidelines and reports. In the talk "Position on Metrics for Security in Requirements Engineering" it is argued that similar to code-testing metrics for security, equivalent requirements level metrics should be defined. Such requirements-level security metrics could then be used in evaluating the quality of software security early on, in order to ensure that the resultant software system possesses the required security characteristics and quality. In the final panel discussion several interesting aspects of quality requirements are highlighted: it is stated that there is a trend towards automation in quality requirements; architects should be integrated in quality requirements engineering and testing; there is a shade of grey in quality requirements, which also affects testing as there is some fuzziness of test output like in performance testing or of test input like in security penetration testing.

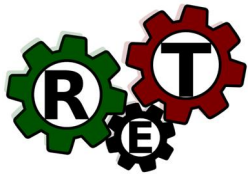

The third paper presentation session comprised three papers on formal languages and models. In first talk "Towards the Automated Generation of Abstract Test Cases from Requirements Models" a model-driven testing approach for conceptual schemas that automatically generates a set of abstract test cases from requirements models is presented. With this approach tests and requirements are linked together to find defects as soon as possible, which can considerably reduce the risk of defects and project reworking. The talk "The Observer-Based Technique for Requirements Validation in Embedded Real-Time Systems" presents an observer-based technique, i.e., a lightweight model-based validation technique to discovering hidden flaws in requirements in the early phases of systems development life cycle. The applicability of the approach is shown in an industrial Vehicle Locking-Unlocking system. The last talk "C&L: Generating Model Based Test Cases from Natural Language Requirements Descriptions" a tool for automatically generating test cases based on Natural Language (NL) requirements specifications is presented. It is shown that the tool is easy to use and at the same time it decreases the time and the effort with respect to test case generation.

Finally, the mapping exercise aiming at jointly defining the area of RET was performed. The exercise and the resulting map of requirements engineering and testing is presented in the next section.

# 4  Map of the Area of RE and Testing

In the afternoon we performed an interactive exercise with the aim of shaping and mapping the area of requirements engineering and testing. The participants were divided into 3 groups consisting of a mix of people from academia and industry. Each group then had 60 minutes to brainstorm around which topics and areas that are covered by RET. The input to the brainstorming session consisted of the list of topics in the Call for paper (Cfp), the presented workshop papers and the participants own experience and insight. Through a combination of joint discussions and individual reflections each group produced a map consisting of a number of topics or areas. Each topic, or area, was noted on a post-it and placed and clustered on a large sheet of butcher paper. The participants were also asked to discuss which topics or areas that which are of particular interest to industry and for which areas more research is needed. Each group briefly presented their findings at the end of the session.

The mapping session was very successful and generated a lot of discussions and interaction between the participants. In total the exercise yielded 70 individual post-it notes each representing a RET topic. The full set of elicited topics can be found in the appendix.

After the workshop three of the organizers (Bjarnason, Felderer and Borg) compiled the outcome of the exercise into one joint map of the area of requirements engineering and testing. This was done by going through the post-its notes one-by-one, discussing and placing them on the final map. This included merging some topics, splitting others and clustering them into the following set of 20 main topics or areas for RET (in no particular order):

- "Good enough" RET, i.e. balancing risk. For example, what traces are essential, what are overhead?
- Approaches that integrate RE and Testing, e.g. BDD
- Assessing and monitoring RET alignment, e.g. with metrics, visualisation (F)
- Automation techniques that support RET, e.g. automatic validation and verification of requirements, use of NL& IR techniques (I)
- Collaboration and communication between RE and Testing roles
- Apply results from other related fields and phenomena, e.g. from non-SW organisations

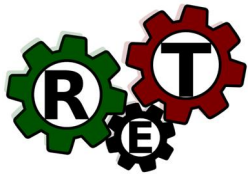

- Context-specific RET, e.g. for OSS development, continuous deployment, safety-critical development, global software development (GSD), non-GSD (I, F)
- Empirical research for RET: applicable and driven by industry needs. For example, empirical studies on the benefits of using MBT (I)
- Maintaining RET and managing change including requirements-based regression testing techniques (I)
- Model-based testing, e.g. MBT at higher abstraction levels, generating MB tcs from NL requirements descriptions (I, F)
- Organizational aspects of RET
- Processes and practices for supporting RET including training and education (I, F)
- Requirements quality impact on testing quality (I, F)
- Requirements-based test techniques, e.g. optimizing testing according to requirements prioritization (I)
- RET at different abstraction levels (I, F)
- RET for large amounts of requirements and test cases
- RET for quality requirements (I, F)
- Test-driven requirements engineering, e.g. the use of A/B testing for requirements prioritization and validation
- Tool support for RET (F)
- Traceability between RE and Testing

The outcome of this mapping exercise (see appendix) provides pointers to topics and areas relevant to RET. The topics which are of particular interest to industry (I) and for which further research is needed (F) were briefly discussed by the groups (indicated in the list above). However these two points need further discussions and investigations in order to provide a clearer roadmap of existing and missing research. After the workshop the participants were invited to suggest key references for the identified topics and to complement the map with potentially missing topics. This initial set of key references is also available in the appendix.

# 5  Future

Due to the interest and positive response to this year's RET workshop both from participants and PC members, we plan to organise the workshop again next year. Our aim is to organise RET'15 co-located with ICSE'15 (http://2015.icse-conferences.org/ ) and thereby solicit participants from both the RE and the Testing community. If the workshop is accepted the date for paper submissions will be January 23, 2015, with author notification on February 18, 2015.

# 6  Acknowledgements

We want to thank the participants of the workshop and all the authors of submitted papers for their important contribution to the event. In addition, we want to thank the organizers of the IEEE International Requirements Engineering Conference (RE 2014), our sponsor EASE - The Industrial Excellence Centre for Embedded Applications Software Engineering (http://ease.cs.lth.se/om/) and the members of the program committee listed below:

**Armin Beer**, Beer Test Consulting, Austria
**Pether Camitz**, System Verification, Sweden
**Nelly Condori-Fernandez**, PROS Research Centre, Spain
**Jane Cleland-Huang**, DePaul University, USA
**Robert Feldt**, Blekinge Institute of Technology, Sweden

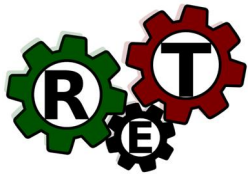

**Vahid Garousi**, University of Calgary, Canada
**Joel Greenyer**, University of Hannover, Germany
**Mats Heimdahl**, University of Minnesota, USA
**Andrea Herrmann**, Herrmann & Ehrlich, Germany
**Jacob Larsson**, Capgemini, Sweden
**Per Lenberg**, SAAB ATM Sensis AB, Sweden
**Emmanuel Letier**, University College London, UK
**Annabella Loconsole**, Malmö University, Sweden.
**Cu Duy Nguyen**, University of Luxembourg, Luxembourg
**Magnus C. Ohlsson**, System Verification, Sweden
**Barbara Paech**, University of Heidelberg, Germany
**Anna Perini**, FBK, Trento, Italy
**Dietmar Pfahl**, University of Tartu, Estonia
**Giedre Sabaliauskaite**, Singapore University of Technology and Design, Singapore
**Kristian Sandahl**, Linköping University, Sweden
**Hema Srikanth**, IBM, USA
**Paolo Tonella**, Fondazione Bruno Kessler, Italy
**Michael Whalen**, University of Minnesota, USA
**Magnus Wilson**, Ericsson, Sweden
**Krzysztof Wnuk**, Lund University, Sweden

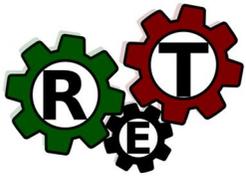

# APPENDIX: Map of the area of RET – result of workshop exercise

| Main area | Post-it | Sub area | Comment | Future research | Industry need | Cluster # | Group # | Key references |
|---|---|---|---|---|---|---|---|---|
| "Good enough" RET | Good enough RET? | | | | | 6 | 2 | |
| | Managing risk | 1 | | | | 6 | 2 | |
| | Requirements and code coverage, objective and good exit criteria | | | | | 6 | 2 | |
| | Good enough traceability, e.g. what traces are essential/what traces are just overhead | 1 | | | | 33 | 1 | |
| Apply results from other related fields and phenomena, e.g. from non-SW organisations | Compare RET alignment with related phenomena in non-SW organisations | | processes, people etc of other fields. E.g. information management | | | 1 | 3 | |
| Approaches that integrate RE and Testing, e.g. BDD | Co-elicitation of requirements and tests | | Draft | | | 19 | 2 | |
| Assessing and monitoring RET alignment, e.g. with metrics, visualisation | Visualisation of RET | | | 1 | | 2 | 3 | |
| | Measuring test case coverage against reqts | | | | | 6 | 2 | |
| | Monitor degree of RET alignment | | | | | 12 | 2 | |
| Automation techniques that support RET | Requirements and test automation | | | | 1 | 5 | 3 | |
| | NL-driven RET (NL technology) | | | | | 9 | 2 | |
| | Automation of requirements validation | | | | | 31 | 1 | |
| | Automation of requirements verification | | | | 1 | 31 | 1 | |
| Collaboration and communication between RE and Testing roles | Knowledge management and transfer between RE and testers | | | | | 4 | 3 | |
| | Team work | | | | | 4 | 3 | |
| | RE-Test communication | | | | | 29 | 1 | |

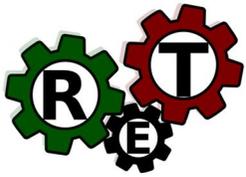

# APPENDIX: Map of the area of RET – result of workshop exercise

| Main area | Post-it | Sub area | Comment | Future research | Industry need | Cluster # | Group # | Key references |
|---|---|---|---|---|---|---|---|---|
| Context-specific RET | Domain- and context-specific RET alignment | | Different practices and challenges depending on domain and context | 1 | | 1 | 3 | |
| | OSS RET alignment | 1 | | | | 1 | 3 | |
| | Continuous Deployment | 1 | | 1 | | 5 | 3 | Clap et al. "On the journey to continuous deployment: Technical and social challenges along the way", IST, vol 57, 2015 |
| | Continuous Integration. Requirements-based test cases. When to test what? | | | | 1 | 5 | 3 | |
| | Certification of safety critical SW | 1 | | | | 6 | 2 | |
| | Global optimization of RET effort | 1 | | | | 15 | 2 | |
| Empirical research for RET: applicable and driven by industry needs | Evidence, Success stories, Empirical studies in the field | | | | 1 | 1 | 3 | |
| | Empirical studies of benefits of using MBT | 1 | | | 1 | 1 | 3 | |
| | Empirical evidence for decisions in RET, e.g. A/B testing | 1 | | | | 6 | 2 | |
| | Empirical studies of co-locating RE and V&V | 1 | | | | 11 | 2 | |
| | More concrete applications of academic methods / concepts | 1 | | | 1 | 22 | 2 | |
| | "old" references: black-box testing | | | | | 25 | 2 | |

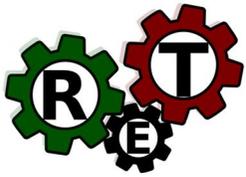

# APPENDIX: Map of the area of RET – result of workshop exercise

| Main area | Post-it | Sub area | Comment | Future research | Industry need | Cluster # | Group # | Key references |
|---|---|---|---|---|---|---|---|---|
| Maintaining RET and managing change incl reqts-based regression testing techniques | Traceability, impact analysis for selecting regression test cases with changing requirements | 1 | | | 1 | 2 | 3 | |
| | Change management in RET | | | | | 20 | 2 | |
| | Evolution according to software evolution | | | | | 21 | 2 | |
| | Test case selection after requirements change | | | | | 30 | 1 | |
| | Maintain traceability over time (continually) | | | | 1 | 33 | 1 | |
| Model-based testing | MBT at the higher requirements levels | 1 | Draft | | | 14 | 2 | |
| | Requirements-based MBT | | | | | 14 | 2 | |
| | Requirements model for automatically generating test cases | | | 1 | | 18 | 2 | |
| | Bi-direction: Reqts model <-> test cases | | | | | 19 | 2 | |
| | "forgotten" oracle problem | | | | | 25 | 2 | |
| | Generating model-based test cases from natural language requirements descriptions | 1 | | | 1 | 31 | 1 | |
| | Model-based approaches to requirements testing | | | | 1 | 31 | 1 | |
| | Domain knowledge (modelling) in bridging gap of RE and Testing | 1 | | 1 | 1 | 35 | 1 | |
| Organisational aspects of RET | Organisational aspects of requirements-test alignment | | | | | 34 | 1 | |

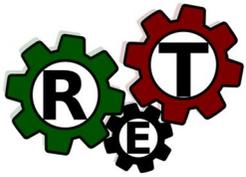

# APPENDIX: Map of the area of RET – result of workshop exercise

| Main area | Post-it | Sub area | Comment | Future research | Industry need | Cluster # | Group # | Key references |
|---|---|---|---|---|---|---|---|---|
| Processes and practices for supporting RET incl training and education | Practices for aligning RE and Testing | | | | 1 | 16 | 2 | |
| | RET training and education | 1 | | | | 23 | 2 | |
| | RET body of knowledge | | | | | 24 | 2 | |
| | Requirements engineers learning testing | | | | | 26 | 1 | |
| | How architecture is involved in requirements | 1 | | | 1 | 27 | 1 | |
| | Workshops on writing requirements | | | | | 28 | 1 | |
| | Process and practices for aligning RE and Testing | | | | 1 | 35 | 1 | |
| | Standards for RE - Test processes. What is the state of the art? | | | 1 | | 35 | 1 | |
| | Standardized interface between RE and Testing teams | | | 1 | | 35 | 1 | |
| | Independent testing vs knowledgeable testers | | | 1 | | 36 | 1 | |
| Requirements quality impact on testing quality | Investigate how RE quality influences testing | | | 1 | | 3 | 3 | |
| | Reqts validation. Reqts completeness & sufficiency | | | | | 7 | 2 | |
| | Requirements testability | | | 1 | 1 | 7 | 2 | |
| Requirements-based test techniques | Reqts-based test techniques | | | | | 8 | 2 | |
| | test case selection | | | | 1 | 10 | 2 | |
| | Optimizing test cases for prioritized requirements | 1 | | | 1 | 13 | 2 | |

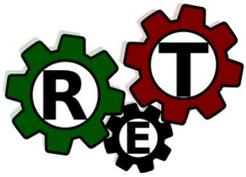

# APPENDIX: Map of the area of RET – result of workshop exercise

| Main area | Post-it | Sub area | Comment | Future research | Industry need | Cluster # | Group # | Key references |
|---|---|---|---|---|---|---|---|---|
| RET at different abstraction levels | Requirements validation and test | | | 1 | | 3 | 3 | |
| | Different levels of testing. Acceptance testing vs system testing. What is the difference? | | | | 1 | 5 | 3 | |
| | V-model with Product mgmt at root: Release criteria (RE) vs Certification (V&V) | | | | | 6 | 2 | |
| | Abstraction levels of requirements | | | | 1 | 31 | 1 | |
| RET for big data | Big data for RET (e.g. A/B testing) | | | | | 6 | 2 | |
| RET for quality requirements | Quality requirements and how to handle test | | | | 1 | 3 | 3 | |
| | Testing non-functional requirements | | | 1 | | 8 | 2 | |
| | co-evolution of quality reqts and tests | | Processes and practices for supporting RET | | | 8 | 2 | |
| Test-driven requirements engineering | Big data for RET (e.g. A/B testing) | | Draft | | | 6 | 2 | |
| Tool support for RET | Tool support for RET alignment | | | 1 | | 2 | 3 | |
| | Tools | | | 1 | | 32 | 1 | |
| Traceability between RE and Testing | Traceability reqts - test | | | | | 17 | 2 | |
| | Traceability btw requirement and test | | | | | 33 | 1 | |